\documentstyle[preprint,aps,psfig]{revtex}
\newcommand{\MeV}{{\rm MeV}}
\newcommand{\fm}{{\rm fm}}
\newcommand{\non}{\nonumber}
\newcommand{\be}{\begin{equation}}
\newcommand{\ee}{\end{equation}}
\newcommand{\bea}{\begin{eqnarray}}
\newcommand{\eea}{\end{eqnarray}}
\begin{document}
\draft
\title{Four-particle condensate in strongly coupled fermion systems}
\author{G.\ R\"opke, A.\ Schnell, P.\ Schuck\cite{isn}}
\address{University of Rostock, FB Physik,
Universit\"atsplatz 1, 18051 Rostock, Germany}
\author{P.\ Nozi\`{e}res}
\address{Institut Laue-Langevin, B.P. 156, 38042 Grenoble Cedex 9,
France}
\date{\today}
\maketitle
\begin{abstract}
Four-particle correlations in fermion systems at finite temperatures
are investigated with special attention to the formation of a
condensate. Instead of the instability of the normal state with
respect to the onset of pairing described by the Gorkov equation, a
new equation is obtained which describes the onset of quartetting.
Within a model calculation for symmetric nuclear matter, we find that
below a critical density, the four-particle condensation ($\alpha$-like
quartetting) is favored over deuteron condensation (triplet pairing).
This pairing-quartetting competition is expected to be
a general feature of interacting fermion systems, such as the
excition-biexciton system in excited semiconductors. Possible
experimental consequences are pointed out.

\noindent Keywords:
Bose-Einstein condensation, superfluidity, strongly coupled systems,
nuclear matter, $\alpha$-particle matter, exciton-biexciton system. 
\end{abstract}

\pacs{05.30.Fk, 03.75.F, 67.60, 21.65}
%%%  Fermion systems and electron gas, 05.30.Fk
%%%  Superfluidity -mixed systems, 67.60
%%%  Bose-Einstein condensation, 03.75.F
%%%  Nuclear matter, 21.65

One of the most amazing phenomena in quantum many-particle systems is
the formation of quantum condensates.
Of particular interest are strongly coupled fermion systems where bound 
states arise. In the low-density limit, where even-number fermionic 
bound states can be considered as bosons, 
Bose-Einstein condensation may be expected to occur at low temperatures.
At present, condensates are investigated in 
systems where the cross-over from Bardeen-Cooper-Schrieffer (BCS) 
pairing to Bose-Einstein condensation (BEC) can be observed, 
see \cite{BEC}. Among very different
quantum systems such as electron-hole exciton system in semiconductors,
atoms in traps at extremely low temperatures, etc., nuclear matter is
particularly suited to study correlation effects in a quantum liquid.

An indication of strong correlations in nuclear matter is the formation of
bound states. The interaction in the singlet ($S=0$) channel is not strong
enough to form a bound state, whereas in the neutron (n) -- proton (p)
triplet ($S=1$) channel a bound
state, the deuteron, arises. In nuclear matter, with increasing density the
bound states are modified and will disappear at the so-called Mott density
\cite{RMS82}. In particular, treating the two-particle Green function in
ladder Hartree-Fock approximation, an effective wave equation (in matrix
notation) $\psi_{\lambda}=K_2(E_{\lambda})\,\psi_{\lambda}$ for the quantum
state $\lambda$ can be derived. Explicitly this reads
\be
\psi_{\lambda}(12)=\sum_{1'2'}K_2(12,1'2',E_{\lambda})\,\psi_{\lambda}(1'2')
\label{two_wave}
\ee
with
\be
K_2(12,1'2',z) = V(12,1'2')\frac{1-f(1)-f(2)}{z-E(1)-E(2)}\:.
\label{two_kernel}
\ee
The influence of the medium is contained in the
single-particle energy $E(1)=p_1^2/2m+\sum_{2}V(12,12)_{\rm ex}f(2)$
and in the Pauli blocking term $[1-f(1)-f(2)]$.
Here $f(1)=[\exp\{E(1)/T-\mu/T\}+1]^{-1}$ is the Fermi distribution
function and '1' stands for momentum, spin, and isospin coordinates,
whereas $V(12,1'2')$ is the antisymmetrized matrix element of the
two--body interaction.

Including medium modifications of two-particle states, a generalized
Beth-Uhlenbeck formula
$N/\Omega_0=n_1(T,\mu)+n_{\rm corr}(T,\mu)$ has been derived
\cite{NSR,SRS}, where the uncorrelated density
$n_1(T,\mu)=\Omega_0^{-1}\sum_1f(1)$ is given by the quasi-particle
contribution ($\Omega_0$ being the normalization volume). The correlated
density $n_{\rm corr}(T,\mu)$ is obtained from the two-particle T matrix.
This approach has been widely applied to ionic plasmas as well as to the
electron-hole exciton system in excited semiconductors \cite{Zimmer}.
Taking into account the Mott effect, general features of the composition
of nuclear matter as function of density and temperature are given
in \cite{ZPhys95}.

At low temperatures it is well known that in nuclei, nuclear matter and
neutron matter (neutron stars) superfluidy can arise in the singlet channel
\cite{BM,Nazarew,ShapT}. A theoretical description of this
superfluidity can be achieved by treating the Gorkov equation
$\psi_2=K_2(\mu_1+\mu_2)\,\psi_2$ for the critical temperature
$T^c_2$ as a function of the chemical potential. It allows the
determination of $T_s^c$ or $T_t^c$ for the singlet and triplet channels,
respectively.

The solution of the Gorkov equation has been considered by different 
authors using realistic nucleon-nucleon interactions. It has been found
that in comparison  with the singlet channel, in the triplet channel
the transition to superfluidity  should arise at relatively high
temperatures \cite{AFRS,BLS}. This is a consequence of the stronger
interaction in the triplet channel which leads to the formation of
a bound state (deuteron) in the low-density limit where $f \ll 1$. 
Estimates give a value of the critical temperature up to
$T_t^c\approx 5\,\MeV$ at one third of the nuclear matter density.
At the same time, at zero temperature a large gap arises \cite{BLS}.
An interesting feature of the triplet pairing in symmetric nuclear
matter is the cross-over from Bose-Einstein condensation of deuterons
at low densities to BCS neutron-proton pairing at high densities
\cite{ZPhys95}. 

In spite of the relatively strong interaction, cf.\ also calculations
with effective pairing forces by Goodman \cite{goodman}, 
triplet pairing seems less apparent in nuclear structure
systematics \cite{Sch94}. However, it should become important for heavier
$N=Z$ nuclei produced in the new radioactive beam facilities.

In this letter we show that in the low-density region the transition
to triplet pairing is not realized, because four-particle
correlations are more dominant there. Obviously, at chemical equilibrium,
in the low-density region at low temperatures the dominant part of nuclear
matter will be found in $\alpha$-particles which are much stronger bound
than the deuteron. Therefore, the triplet pairing (Bose condensation of
deuterons) has to compete with quartetting (Bose condensation of
$\alpha$-particles).

The four-particle correlations are obtained from the four-particle Green
function (compare \cite{DanSch}) that is given in ladder
Hartree-Fock approximation by the equation 
\begin{eqnarray}\label{g4_full}
&&\!\!\!G_4(1234,1'2'3'4',z) = \frac{f(1)f(2)f(3)f(4)}{g_4(1234)}
\,\frac{\delta_{11'}\delta_{22'}\delta_{33'}\delta_{44'}}
{z-E_4(1234)}\non\\
+&&\!\!\!\!\sum_{1''2''3''4''}\!\!\!\!\!K_4(1234,1''2''3''4'',z)
G_4(1''2''3''4'',1'2'3'4',z)\,,\!\!
\end{eqnarray}
where we use the abbreviation $E_n(12\dots n)=E(1)+E(2)+\cdots+E(n)$, and
$g_n(12\dots n) = [\exp(E_n(12\dots n)-n\mu)/T-1]^{-1}$ being the Bose
distribution function.

The instantaneous part of interaction kernel is obtained by using the
technique of Matsubara Green functions 
as
\begin{eqnarray}\label{k4_full}
K_4(1234,1'2'3'4',z) &=& V(12,1'2')
\frac{f(1)f(2)}{g_2(12)} \frac{\delta_{33'}\delta_{44'}}{z-E_4(1234)}
\nonumber\\  &+&\,{\rm perm.}\,,
\end{eqnarray}
where the terms obtained by renumbering are not given explicitly.
We have used the identity $\bar{f}(1) \bar{f}(2) \cdots \bar{f}(n)
-f(1) f(2) \cdots f(n) = g_n^{-1}(12\dots n) f(1) f(2) \cdots f(n)$
with $\bar{f}=1-f$. 

Near a pole the Green function $G_4$ can be factorized, $G_4(1234,1'2'3'4',z) 
\approx\psi_{\nu}(1234)\psi^*_{\nu}(1'2'3'4')/(z-E_\nu)$.
The eigenvalues $E_{\nu}$ and eigenstates $\psi_{\nu}(1234)$ follow
from the solution of the four-particle Schr\"odinger-like wave equation
$\psi_{\nu}=K_4(E_{\nu})\,\psi_{\nu}$. The eigenvalue $E_0(T,\mu)$ of the
lowest bound state ($\alpha$-particle) depends on temperature $T$ and chemical
potential $\mu$ and also on its center-of-mass momentum due to the
medium-dependent self-energy as well as the phase space occupation factors.

From the four-particle Green function, the four-particle density
matrix is obtained as
\begin{eqnarray}
\label{4dens}
&&<a_1^+a_2^+a_3^+a_4^+a_{4'}a_{3'}a_{2'}a_{1'}>\non\\
&& =\int {d\omega \over \pi
} g_4(\omega){\rm Im} G_4(1234,1'2'3'4',\omega-i0)\,.
\end{eqnarray}
Obviously, the four-particle density (\ref{4dens}) diverges when, at a given
temperature $T$, the chemical potential takes the value  $\mu_c=E_0
(T,\mu_c)/4$. Then, the delta function produced by the pole of $G_4$
coincides with the singularity of the Bose function. 
Values for the chemical potential $\mu$ exceeding the value of the
lowest bound state energy $E_0(T,\mu)$ are not admissible because the
diagonal elements of the four-particle density are positive
definite.

The main objective of this letter is to give an estimate of the critical
temperature $T_4^c(\mu)$ for the onset of a four-particle condensate. This
is obtained by solving the equation
\be
\psi_4(1234)=\sum_{1'2'3'4'}K_4(1234,1'2'3'4',4\mu)\psi_4(1'2'3'4')\,.
\label{tc}
\ee
In the low-density limit, where the distribution functions occurring in $K_4$
are small compared with 1, this equation coincides with the Schr\"odinger
equation for the four-particle bound state, i.e.\ the $\alpha$-particle
in free space.

To discuss the competition between two-particle pairing and the condensation
of four-particle states, we perform an exploratory calculation for a simple
model system which contains the formation of a two-particle bound state in the
triplet channel and a singlet four-particle bound state. We use a separable
interaction of Gaussian form $V_{s/t}(12,1'2')=-\lambda_{s/t}\Omega_0^{-1}
\exp(-p_{12}^2/b^2)\,\exp(-p_{12}'^2/b^2)\,\delta(q_{12}-q_{12}')$
with the relative momentum $p_{12}=(p_1-p_2)/2$ and the center-of-mass momentum
$q_{12}=p_1+p_2$. At given range parameter $b=1.54\,\fm^{-1}$, the interaction
strengths are adopted as
$\lambda_t=1213.8\,\MeV\fm^3$ and $\lambda_s=536\,\MeV\fm^3$ to reproduce the
free deuteron binding energy $E_t=-2.225\,\MeV$ in the neutron-proton
triplet channel as 
well as the free $\alpha$ binding energy $E_0=-28.29\,\MeV$ for the
four-particle ground state (see below). In principle it is possible to extend
the calculations to more realistic nucleon-nucleon potentials as given
in \cite{pot}.

\begin{figure}[htb]
\centerline{\psfig{figure=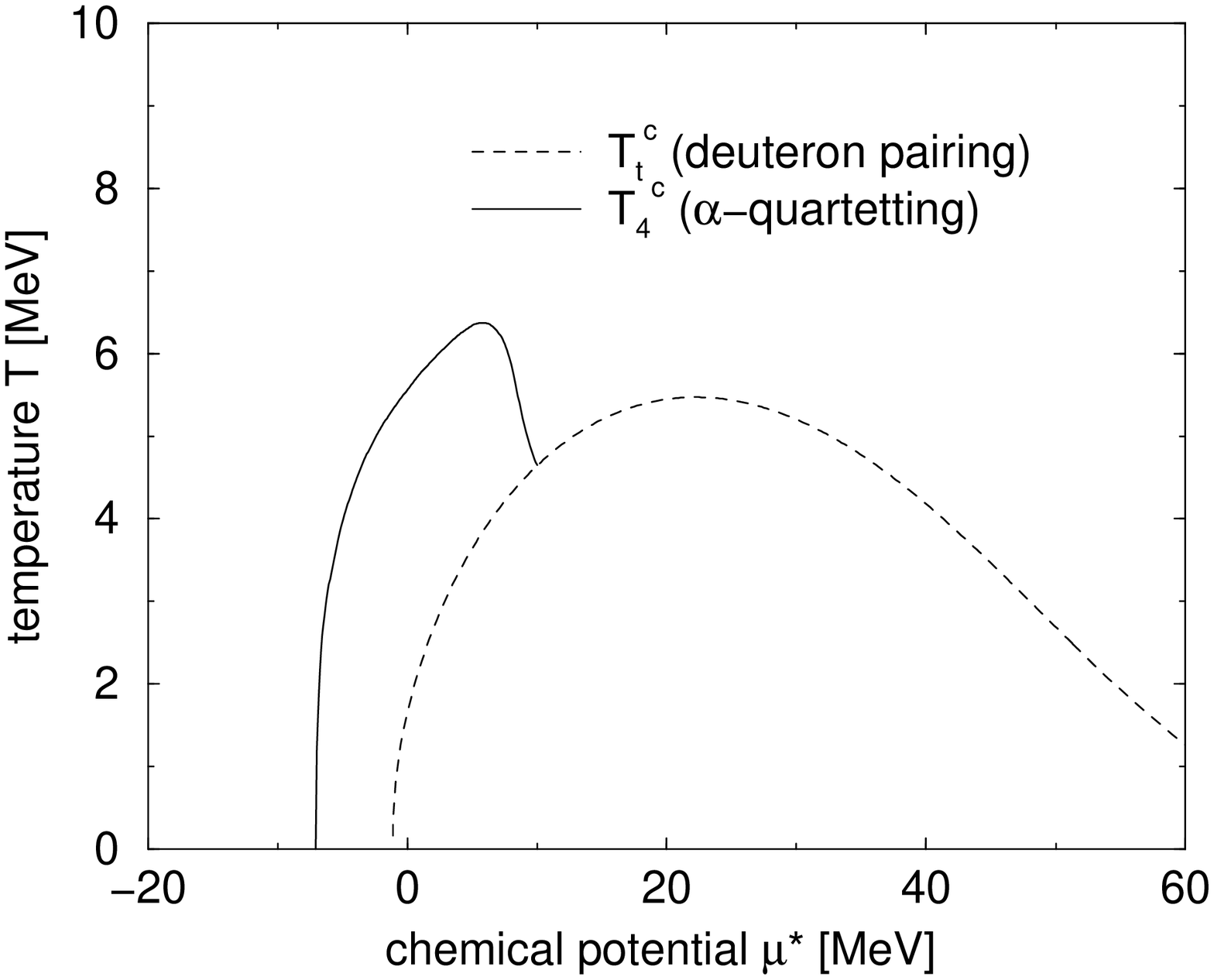,width=7.5cm}}
\caption{Critical temperatures for the onset of quantum condensation in
symmetric nuclear matter, model calculation. The critical temperature of the
onset of two-particle pairing $T_t^c$ is compared with $T_4^c$ for the onset
of a four-particle condensate, as a function of the chemical potential.}
\label{fig_tc_mue}
\end{figure}

\begin{figure}[htb]
\centerline{\psfig{figure=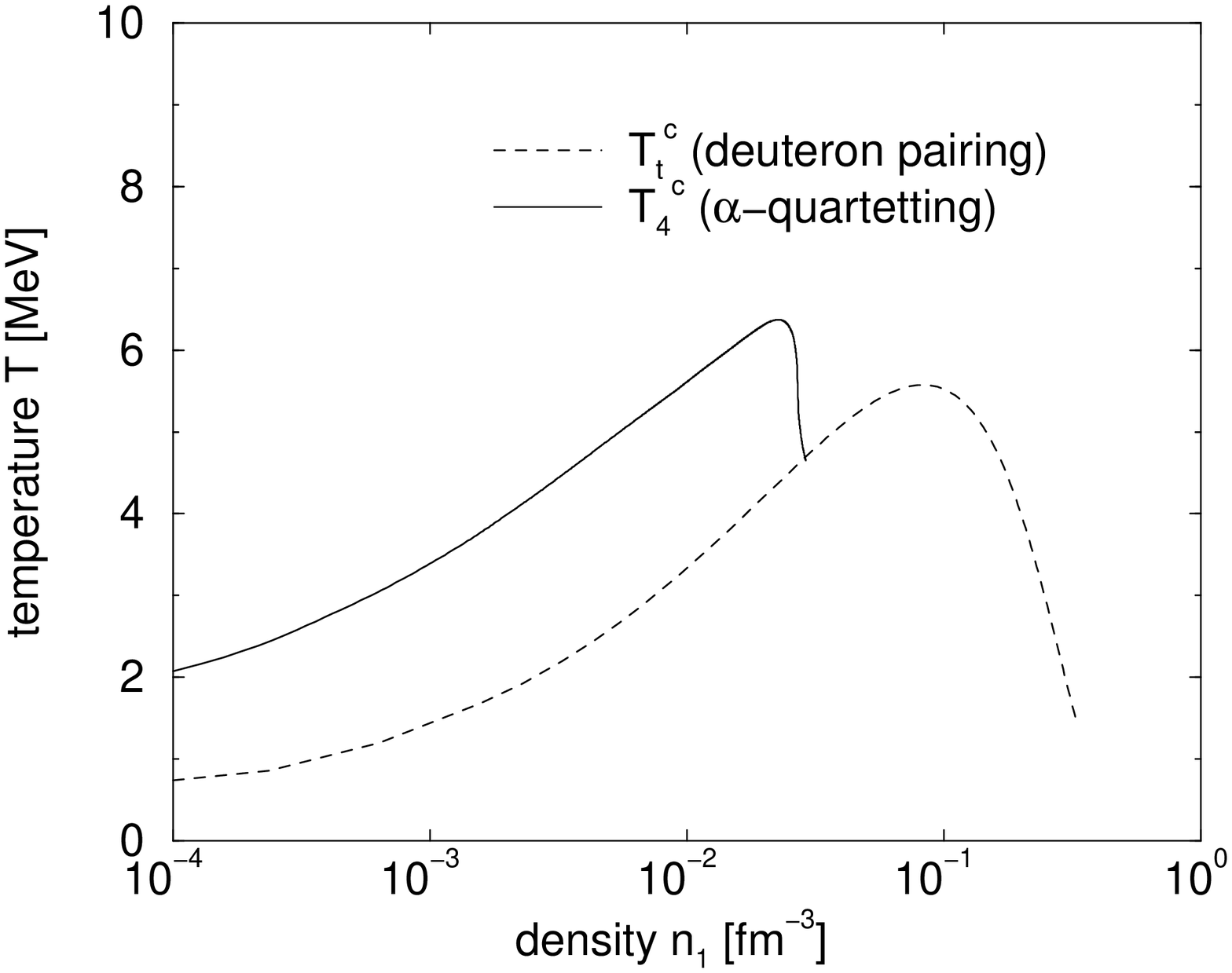,width=7.5cm}}
\caption{The same as Fig.~\protect\ref{fig_tc_mue} but as a function of the
uncorrelated density $n_1$.}
\label{fig_tc_rho}
\end{figure}

We begin with the discussion of neutron-proton triplet pairing. The critical
temperature $T_t^c$ obtained from the solution of the Gorkov equation is
presented as a function of the reduced chemical potential $\mu^*$ in
Fig.~\ref{fig_tc_mue} and as a function of the uncorrelated density
$n_1(T,\mu)=\Omega_0^{-1}\sum_1f(1)$ in Fig.~\ref{fig_tc_rho} (dotted lines). 
For the sake of simplicity the self-energy shift is taken to be a constant
that is incorporated into a shift of the chemical potential,
$\mu^*=\mu-\sum_2V(12,12)_{\rm ex}f(2)$, at $p_1=0$. 

The four-particle wave equation (\ref{g4_full}) is solved within a variational
ansatz where the four-particle wave function with zero total momentum is given
as a product of two wave functions for the relative motion (quasi deuterons)
$\phi(p)$ and the center-of-mass wave function $\psi(q)$,
\be
\psi_4(1234) = \phi(p_{12})\phi(p_{34})\,\psi(q_{12})
\delta_{q_{12}+q_{34},0}\,.\label{ansatz}
\ee
The wave function for the relative motion $\phi(p)$ is optimized in a space
of functions containing the exact solution for the wave function of the
in-medium deuteron (Eqs.~(\ref{two_wave},\ref{two_kernel})), whereas for the
center-of-mass wave function $\psi(q)$ a simple Gaussian form is adopted.

The results for the critical temperature $T_4^c$ as a function of the chemical
potential $\mu^*$ and the uncorrelated density $n$ are displayed in
Figs.~\ref{fig_tc_mue} and \ref{fig_tc_rho}, respectively (solid lines).
In the low-temperature limit, with increasing chemical potential the transition
to quartetting occurs prior to the pairing transition. This is a consequence of
the fact that the value $E_0/4$ for the four-particle bound state lies below
the value $E_t/2$ of the triplet bound state. The Gorkov equation, which
predicts a normal state for temperatures $T>T_t^c$, is not applicable in that
region, because already at a higher temperature $T_4^c$ the normal state was
removed due to the onset of quartetting.

A simple argument for the behavior of $T_4^c$, $T_t^c$ as a function of $n_1$
in the limit of low densities can be given
from the law of mass action. Neglecting medium corrections the critical
temperatures $T_t^c$ and $T_4^c$ as functions of $n_1$ are found as solutions
of $n_1=4(mT^c_t/2\pi\hbar^2)^{3/2}\exp\{E_t/2T^c_t\}$ and
$n_1=4(mT^c_4/2\pi\hbar^2)^{3/2}\exp\{E_0/4T^c_4\}$, respectively.

With increasing density the critical temperature $T_4^c$ approaches $T_t^c$
from above at a critical density $n_1^0=0.03 fm^{-3}$
($\mu^*_0\approx 10\,\MeV$).
For $n_1<n_1^0$ ($\mu^*<\mu^*_0$) the four-particle condensate
arises whereas for
$n_1>n_1^0$ the system goes over to a BCS pairing state.

The fact that at higher densities $\alpha$-particle condensation disappears
faster than deuteron Cooper pairing is at first sight quite astonishing
because of the very strong binding of $\alpha$'s in free space. This
feature, one of the main results of our work, can, however, be explained
qualitatively with simple arguments and is generic for any Fermi system
where pairing and quartetting may interfere.
Indeed the competition between pair and quartet condensation in
attractive Fermi liquids is known also from the field of semiconductors,
where both exciton and biexciton condensates have been suggested.
The qualitative features of that competition are easily understood.
Since they do not depend crucially on the triplet nature of the condensate we
discuss the simpler case of singlet pairing as a function of density.
In the weak coupling limit (high density), 
the ground state is described by the standard
BCS wave function. However weak the attraction is, a pairing amplitude
$\langle aa\rangle$ appears due to the finite density of states at Fermi
level for pairs with zero total momentum $q$ (the two-particle kernel
has a logarithmic singularity).
In contrast a quartet condensate $\langle aaaa\rangle\neq 0$
does not exist on its own. Due to the exclusion principle the corresponding
density of states for $q=0$ vanishes at Fermi level and it takes a minimum
attraction to develop a Cooper pole.

In the opposite limit of dilute systems with attraction strong enough to bind
individual pairs and quartets, an atomic regime prevails. Single pairs
$\langle aa\rangle$ may be viewed as bosons $\langle\Phi\rangle$ and the
quartet as a bound pair of two bosons. The competition between particle and
pair condensation in Bose liquids was discussed in \cite{NSJ}.
In the dilute limit we expect ``molecular'' Bose condensation of bound bosons,
as described long ago by Valatin and Butler \cite{VB}. In the dense limit the
quartets dissociate and single boson condensation takes over. As the density
decreases, a variational Ansatz \cite{NSJ} shows that the condensate fraction
$n_0=|\langle\Phi\rangle|^2$ goes down, vanishing at a critical $n_c$ beyond
which only $\langle\Phi\Phi\rangle$ survives ($n_c$ also marks the appearance
of a gap).

Altogether, we expect a quartet atomic condensate for very low densities and
a BCS pair condensate for very high densities. In between the interpolation
depends on the pair and quartet binding energies, $E_2$ and
$2E_2+E_4$, as compared to the Fermi energy. If
$E_2\gg E_4$ the pair condensate enters the atomic regime well
before it disappears in favor of quartets. If on the other hand
$E_2<E_4$ the system goes directly from a BCS pair superfluid
to a quartet Bose-Einstein condensation.

Possible signatures of quartetting have been discussed in the past in the
context of nuclear structure \cite{Cau81}. According to the present
studies quartet correlations may be of importance in the outer
regions of nuclei. We can make a rough estimate for the critical
temperature of $\alpha$ - particle condensation in nuclei in
converting $T_4^c(n)$ of Fig. 2 via the Local Density Approximation
into a radius dependence $T_4^c(r)$ and averaging over a typical
nuclear density. This procedure applied to ordinary neutron--neutron
pairing yields $T_c \approx $ 1 MeV in good agreement with more
microscopic calculations \cite{ST}. In this way we predict that the
increase of the critical temperature due to $\alpha$ - particle
condensation is about $0.2\;\MeV$.
In spite of our limited variational approach, we think that this value
should be correct within 30 to 40\%. Also the critical density
$n_1^0=0.03\,\fm^{-3}$ for the onset of quartetting should be reliable
within these limits and therefore our estimate can be of interest for
clustering phenomena in stellar collapse and the outer crust of neutron stars
\cite{ShapT}.
We believe that in the far tail of nuclear densities ($n<n_1^0$) quartetting
can occur as well as in a number of nuclear processes such as
the emission of $\alpha$-particles from the heaviest elements
\cite{Merchant} and unusually 
large $\alpha$-decay rates of neutron deficient lead isotopes \cite{Dum94}.
Bose-Einstein effects may also be of relevance in heavy-ion collisions at a
relatively late stage where the temperature of expanding matter has dropped
below $5\,\MeV$.

In conclusion, we have shown that prior to the transition to pairing, a
transition from the normal state to four-particle condensation according to
Eq.~(\ref{tc}) can occur. However, as a generic phenomenon 4-body
condensation always looses against weak coupling Cooper pairing at
sufficiently high densities.
For demonstration purposes we have considered nuclear matter as a
strongly coupled quantum liquid. The possibility of quartetting and higher
correlated condensates is of interest also in other systems, e.g.,
electron-hole pairs in semiconductors forming bi-excitons \cite{BEC,IK}
or quarks forming two-pion states \cite{ACSW}.

G.\ R.\ and A.\ S.\ appreciate useful discussions with G. Bertsch.
P.\ S.\ is grateful to Deutsche Forschungsgemeinschaft (DFG) for financial
support and to University of Rostock for pleasant hospitality.


\begin{references}
\bibitem[*]{isn}
Permanent Address: Institut de Sciences Nucl\'{e}aires,
Universit\'{e} Joseph Fourier, CNRS-IN2P3
53, Avenue des Martyrs, F-38026 Grenoble Cedex, France
\bibitem{BEC}
A. Griffin, D. W. Snoke, and S. Stringari (eds.),
{\it Bose-Einstein Condensation} (Cambridge University Press,
Cambridge, 1995).
\bibitem{RMS82}
G. R\"opke, L. M\"unchow, and H. Schulz,
Nucl. Phys. {\bf A379}, 536 (1982); Phys. Lett. B {\bf 110}, 21 (1982);
G. R\"opke, M. Schmidt, L. M\"unchow, and H. Schulz,
Nucl. Phys. {\bf A399}, 587 (1982).
\bibitem{NSR}
P. Nozi\`eres and S. Schmitt-Rink,
J. Low Temp. Phys. {\bf 59}, 159 (1985).
\bibitem{SRS}
M. Schmidt, G. R\"opke, and H. Schulz,
Ann. Phys. (NY) {\bf 202}, 57 (1990).
\bibitem{Zimmer}
R. Zimmermann,
{\it Many-Particle Theory of Highly Excited Semiconductors}
(Teubner, Leipzig, 1988).
\bibitem{ZPhys95}
H. Stein, A. Schnell, T. Alm, and G. R\"opke,
Z. Phys. A {\bf 351}, 295 (1995).
\bibitem{BM}
A. Bohr and B. R. Mottelson
{\it Nuclear Structure} (Benjamin, N.Y., 1969), Vol. I;
P. Ring and P. Schuck,
{\it The Nuclear Many-Body Problem} (Springer, N.Y., 1980).
\bibitem{Nazarew}
J. Dobaczewski, W. Nazarewicz, T. R. Werner, J. F. Berger, C. R. Chinn,
and J. Decharg\'{e},
Phys. Rev. C {\bf 53}, 2809 (1996).
\bibitem{ShapT}
S. L. Shapiro and S. A. Teukolsky, {\it Black Holes, White Dwarfs and Neutron
Stars: The Physics of Compact Objects} (Wiley, N.Y., 1983);
D. Pines, R. Tamagaki, and S. Tsurata (eds.), {\it Neutron Stars}
(Addison-Wesley, N.Y., 1992).
\bibitem{AFRS}
T. Alm, G. R\"opke, M. Schmidt,
Z. Phys. A {\bf 337}, 355 (1990);
T. Alm, B. L. Friman, G. R\"opke, and H. Schulz,
Nucl. Phys. {\bf A551}, 45 (1993);
B. E. Vonderfecht, C. C. Gearhart, W. H. Dickhoff, A. Polls, and A. Ramos,
Phys. Lett. B {\bf 253}, 1 (1992);
M. Baldo, I. Bombacci, and U. Lombardo,
Phys. Lett. B {\bf 283}, 8 (1993).
\bibitem{BLS}
M. Baldo, U. Lombardo, and P. Schuck,
Phys. Rev. C {\bf 52}, 975 (1995).
\bibitem{goodman}
A. L. Goodman,
Nucl. Phys. {\bf A352}, 30 and 45 (1981);
Nucl. Phys. {\bf A369}, 365 (1981).
\bibitem{Sch94}
M. Baldo, U. Lombardo, and P. Schuck,
Phys. Rep. {\bf 242}, 159 (1994).
\bibitem{DanSch}
P. Danielewicz and P. Schuck,
Nucl. Phys. {\bf A567}, 78 (1994);
S. Koh,
Phys. Rev B {\bf 49}, 8983 (1994);
J. Dukelsky, G. R\"opke, and P. Schuck,
Nucl. Phys. {\bf A} (to be published).
\bibitem{pot}
V. G. J. Stoks {\it et al.},
Phys. Rev. C {\bf 49}, 2950 (1994);
R. B. Wiringa {\it et al.},
Phys. Rev. C {\bf 51}, 38 (1995);
R. Machleidt, F. Sammarruca, and Y. Song,
Phys. Rev. C {\bf 53}, 1483 (1996).
\bibitem{NSJ}
P. Nozi\`{e}res and D. Saint James,
J. Physique {\bf 43}, 1133 (1982).
\bibitem{VB}
J. G. Valatin and D. Butler,
Nuovo Cimento {\bf 10}, 37 (1958).
\bibitem{Cau81}
M. Cauvin, V. Gillet, and F. Soulmagnon,
Nucl. Phys. {\bf A361}, 192 (1981);
Y. K. Gambhir, P. Ring, and P. Schuck,
Phys. Rev. Lett. {\bf 51}, 1235 (1983);
K. Varga, R. G. Lovas, and R. J. Liotta,
Phys. Rev. Lett. {\bf 69}, 37 (1992);
F. Aldabe, G. G. Dussel, and H. M. Sofia,
Phys. Rev. C {\bf 50}, 1518 (1994).
\bibitem{ST}
P. Schuck and K. Taruishi, Phys. Lett. B {\bf 385}, 12 (1996).
\bibitem{Merchant}
B. Buck, A. C. Merchant, and S. M. Perez,
Phys. Rev. Lett. {\bf 72}, 1326 (1994);
B. Buck, J. C. Johnston, A. C. Merchant, and S. M. Perez,
Phys. Rev. C {\bf 53}, 2841 (1996).
\bibitem{Dum94}
O. Dumitrescu,
Phys. Rev. C {\bf 49}, 1466 (1994).
\bibitem{IK}
A. L. Ivanov, L. V. Keldysh and V. V. Panashchenko, Sov. Phys. JETP
{\bf 72}, 359 (1991).
\bibitem{ACSW}
T. Alm, G. Chanfray, P. Schuck, and G. Welke,
Nucl. Phys. {\bf A612}, 472 (1997).
\end{references}
\end{document}